\documentclass[12pt]{article}
\usepackage{graphics,epsfig}
\usepackage[english]{babel}
\newcommand{\cinst}[2]{$^{\mathrm{#1}}$~#2\par}
\newcommand{\crefi}[1]{$^{\mathrm{#1}}$}

\begin{document}


\thispagestyle{empty}




\begin{center}

\vglue 1.0cm {\Large \textbf{ The yields of light meson
resonances\\  in neutrinonuclear interactions \\
\vspace{0.5cm} at $\langle E_\nu \rangle \approx$ 10 GeV}}

\end{center}

\vspace{1.cm}

\begin{center}
{\large SKAT Collaboration}

 N.M.~Agababyan\crefi{1}, V.V.~Ammosov\crefi{2},
 M.~Atayan\crefi{3},\\
 N.~Grigoryan\crefi{3}, H.~Gulkanyan\crefi{3},
 A.A.~Ivanilov\crefi{2},\\ Zh.~Karamyan\crefi{3},
V.A.~Korotkov\crefi{2}

\setlength{\parskip}{0mm}
\small

\vspace{1.cm} \cinst{1}{Joint Institute for Nuclear Research,
Dubna, Russia} \cinst{2}{Institute for High Energy Physics,
Protvino, Russia} \cinst{3}{Yerevan Physics Institute, Armenia}
\end{center}
\vspace{100mm}

{\centerline{\bf YEREVAN  2008}}


\newpage
\vspace{1.cm}
\begin{abstract}

\noindent The total yields of the all well-established light
mesonic resonances (up to the $\phi$(1020) meson) are estimated in
neutrinonuclear interactions at $\langle E_\nu \rangle \approx$ 10
GeV, using the data obtained with SKAT bubble chamber. For some
resonances, the yields in the forward and backward hemispheres in
the hadronic c.m.s. are also extracted. From the comparison of the
obtained and available higher-energy data, an indication is
obtained that the resonance yields rise almost linearly as a
function of the mean mass $\langle W \rangle$ of the
neutrinoproduced hadronic system. The fractions of pions
originating from the light resonance decays are inferred.

\end{abstract}

\newpage
\setcounter{page}{1}
\section{Introduction}

\noindent The space-time pattern of the leptoproduced quark-string
fragmentation into hadrons would be rather incomplete without
discerning to what extent the hadrons detected in a given
phase-space domain originate directly from the string
fragmentation and to what they are decay products of other,
higher-mass string fragments$-$resonances. Although it is
generally accepted that the later cause a significant fraction of
the yields of stable hadrons (pions and kaons), a quantitative
estimation of this fraction is not available yet, at least for
neutrino-induced reactions. At present, more or less detailed
experimental data on the neutrinoproduction of mesonic resonances
are available for $\rho$ mesons (\cite{ref1,ref2,ref3,ref4,ref5}
and references therein), and for charged $K^*(892)$ mesons
(\cite{ref6,ref7} and references therein), while those for other
resonances are rather scarce and obtained at high energies of
(anti)neutrino, $\langle E_\nu \rangle \sim$ 40-50 GeV
\cite{ref2}. The aim of this work is to measure, at the same
experimental conditions, the yields of the all well-established
light mesonic resonances (with masses up to $\sim$ 1 GeV$/c^2$) in
neutrinonuclear charged current interactions at intermadiate
energies ($\langle E_{\nu}\rangle \sim$ 10 GeV). In Section 2, the
experimental procedure is described. Section 3 presents the
experimental data on the total yields of 12 mesonic resonances:
$\eta$, $\rho^0$, $\rho^+$, $\rho^-$, $\omega$, $K^*(892)^0$,
$\overline{K}^*(892)^0$, $K^*(892)^+$, $K^*(892)^-$,
${\eta}'(958)$, $f_0(980)$ and $\phi$. For same cases, the
differential yields in the forward and backward hemispheres (in
the hadronic c.m.s.) are also presented. The dependence of the
resonance yields on their mass and the invariant mass $W$ of the
created hadronic system is compared to the higher-energy
neutrinoproduction and $e^+e^-$ annihilation data. Section 4 is
devoted to the estimation of the fraction of pions originating
from the decays of the light meson resonances. The results are
summarized in Section 5.

\section{Experimental procedure}

\noindent The experiment was performed with SKAT bubble chamber
\cite{ref8}, exposed to a wideband neutrino beam obtained with a
70 GeV primary protons from the Serpukhov accelerator. The chamber
was filled with a propane-freon mixture containing 87 vol\%
propane ($C_3H_8$) and 13 vol\% freon ($CF_3Br$) with the
percentage of nuclei H:C:F:Br = 67.9:26.8:4.0:1.3 \%. A 20 kG
uniform magnetic field was provided within the operating chamber
volume.
\\ Charged current interactions containing a negative muon with momentum
$p_{\mu} >$0.5 GeV/c were selected. Other negatively charged
particles were considered to be $\pi^-$ mesons, except for the
cases explained below. Protons with momentum below 0.6 GeV$/c$ and
a fraction of protons with momentum 0.6-0.85 GeV$/c$ were
identified by their stopping in the chamber. Non-identified
positively charged particles were considered to be ${\pi}^+$
mesons, except for the cases explained below. Events in which
errors in measuring the momenta of all charged secondaries and
photons were less than 60\% and 100\%, respectively, were
selected. The mean relative error $\langle \Delta p/p \rangle$ in
the momentum measurement for muons, pions and gammas was,
respectively, 3\%,
 6.5\% and 19\%. Each event is given a weight which corrects for
the fraction of events excluded due to improperly reconstruction.
More details concerning the experimental procedure, in particular,
the estimation of the neutrino energy $E_{\nu}$ and the
reconstuction of $\pi^0\rightarrow 2\gamma$ decays can be found in
our previous publications \cite{ref9,ref5}. \\ The events with $3<
E_{\nu} <$ 30 GeV were accepted, provided that the reconstructed
mass $W$ of the hadronic system exceeds 1.8 GeV. No restriction
was imposed on the transfer momentum squared $Q^2$. The number of
accepted events was 5242 (6557 weighted events). The mean values
of the kinematical variables were $\langle E_{\nu} \rangle$ = 9.8
GeV, $\langle W \rangle$ = 2.8 GeV, $\langle W^2 \rangle$ = 8.7
GeV$^2$,
$\langle Q^2 \rangle$ = 2.6 (GeV/c)$^2$. \\
About 8\% of neutrino interactions occur on free hydrogen. This
contribution was subtracted using the method described in
\cite{ref10,ref11}. The effective atomic weight of the composite
nuclear target is estimated \cite{ref9} to be approximately equal
to $A_{eff}$ = 21$\pm$1, thus allowing to compare our results with
those obtained in $\nu(\overline{\nu})-Ne$ interactions at higher
energies
\cite{ref2}.\\
When considering the production of resonances decaying into
charged kaon(s), the $K^-$ and $K^+$ hypothesis was applied,
respectively, for negatively charged particles and non-identified
positively charged particles (provided that the kaon hypothesis is
not rejected by the momentum-range relation in the propane-freon
mixture), introducing thereat proper corrections for the momentum
of these particles.

\section{Experimental results}

a) {\it The experimental mass resolutions and the fitting
procedure of mass distributions}

\noindent The experimental mass resolutions for different
resonances are estimated from Monte-Carlo simulations. The FWHM
values ($\Gamma^R_{exp}$) of simulated distributions are presented
in Table 1.

\begin{table}[ht]
\caption{The experimental mass resolutions (FWHM's) for different
resonances.}
\begin{center}
\begin{tabular}{|l|c|c|c|c|c|c|c|c|c|}
 \hline
resonance&$\rho^0$&$f_0$&$\rho^{\pm}$&$\eta$ &$\omega$ &$\eta'$
&$K^{*0}$&$K^{*\pm}$&$\phi$ \\
 \hline
 decay mode
 &$\pi^+\pi^-$&$\pi^+\pi^-$&$\pi^{\pm}\pi^0$&$\pi^+\pi^-\pi^0$&$\pi^+\pi^-\pi^0$
 &$\pi^+\pi^-\gamma$&$K^{\pm}\pi^{\mp}$&$K_s^0\pi^{\pm}$&$K^+K^-$ \\
 \hline
 $\Gamma^R_{exp}$(MeV)&47&55&110&66&90&100&28&30&10 \\
 \hline
\end{tabular}
\end{center}
\end{table}

\noindent In the cases, when $\Gamma^R_{exp}$ significantly
exceeds the resonance natural width, $\Gamma^R_0$, the mass
distributions were fitted as a sum of the background ($BG(m)$) and
Gaussian ($G_R(m)$) distributions,
\begin{equation}
dN/dm = BG(m)\, + \, \alpha_R \, G_R(m) \, \, ,
\end{equation}

\noindent where for the Gaussian width $\sigma_R$ an approximate
relation was used: $\Gamma^{exp}_R \approx 2
\sigma_R\sqrt{2\ln2}$. \\
Otherwise, the mass distributions were fitted by the form
\begin{equation}
dN/dm = BG(m) \, \cdot \, (1 + \, \alpha_R \, BW_R(m)) \, \, ,
\end{equation}
\noindent where $BW_R(m)$ is the corresponding Breit-Wigner
function \cite{ref12}, with $\Gamma^R_0$ replaced by an effective
$\Gamma^R_{eff}$ estimated from simulations in which the BW
function is smeared taking into account the experimental
resolution. \\
The form (1) was applied for $\eta$, $\omega$, $\eta'$ and $\phi$,
while the form (2) was used for $\rho^0$, $\rho^{\pm}$,
$K^*(892)^0$ and $K^*(892)^{\pm}$.
 For $f_0(980)$, for which $\Gamma^R_{exp}$ does not
much exceed its natural width $\Gamma^f_0$ = 35 MeV (taken from a
recent NOMAD measurement \cite{ref3}), both (1) and (2) forms
were used, leading to compatible results. \\
The pole mass of $f_0(980)$, $m_{f_0}$ = 963 MeV, was also taken
from \cite{ref3}, while for other resonances considered in this
paper the masses and widths are fixed according to the PDG values
\cite{ref13}. \\ In general, the background distribution was
parametrized as
\begin{equation}
BG(m) = \, B\cdot{(m-m_{th})}^{\beta} \, \exp
(\sum^{k}_{i=1}\varepsilon_im^i)  \, \, \, ,
\end{equation}
\noindent where $m_{th}$ is the threshold mass of the
corresponding resonance; k = 1 or 2, depending on the statistics;
B, $\beta$ and $\varepsilon_i(i=1,k)$ are fit parameters. In same
cases, depending on the form of the mass distribution, the
parameter $\beta$ was fixed to 0.

b) {\it Non-strange resonances}

\noindent The ($\pi^+\pi^-$) effective mass distribution is
plotted in Figure 1 (the left panels) for the whole range of the
Feynman $x_F$ variable, as well as for the forward ($x_F>0$) and
backward ($x_F<0$) hemispheres in the hadronic c.m.s. Signals for
$\rho^0$ and $f_0(980)$ production are visible, except for
$f_0(980)$ at $x_F<0$. The corresponding mean multiplicities are
quoted in Table 2 (the data on $f_0(980)$ are corrected for the
$\pi^+\pi^-$ decay fraction). These values are, as expected,
somewhat smaller than those estimated recently \cite{ref4} at a
slightly severe cut on $W$ ($W>2$ instead of $W>1.8$ GeV, or
$\langle W \rangle = 3.0$ instead of $\langle W \rangle = 2.8$
GeV). As it is seen from Table 2, the $\rho^0$ and $f_0(980)$
production occurs predominantly in the
forward hemisphere. \\
The total yields of charged $\rho$ mesons and the differential
yields of $\rho^+$ at $x_F>0$ and $x_F<0$ were measured in our
previous work \cite{ref5} where the same cut $W>1.8$ GeV was
applied as in the present study. In this work, an attempt is
undertaken to estimate the $\rho^-$ yields at $x_F>0$ and $x_F<0$
too. Figure 1 shows the effective mass distributions for
$\pi^+\gamma\gamma$ and $\pi^-\gamma\gamma$ systems. The
distributions are corrected for losses of reconstructed $\pi^0$
and contamination from the background $\gamma\gamma$ combinations
(see \cite{ref5} for details). The $\rho^-$ signal is observable
at $x_F<0$, but not at $x_F>0$, occupied mainly by favorable
mesons which can, unlike $\rho^-$ meson, contain the current
quark. The total and differential yields of $\rho^+$ and $\rho^-$
are quoted in Table
2. \\
The signals for $\eta$ and $\omega$ production were looked for in
decays $\eta\rightarrow\pi^+\pi^-\pi^0$ (with 22.7\% branching
fraction) and $\omega\rightarrow\pi^+\pi^-\pi^0$ (with 89\%
branching fraction). The $\pi^+\pi^-\gamma\gamma$ effective mass
distributions in three $x_F$-ranges are plotted in Figure 2 (the
left panels). As for the case of $\rho^{\pm}$ \cite{ref5}, the
distributions are corrected for losses of reconstructed $\pi^0$
and the background $\gamma\gamma$ contamination. It is seen from
Figure 2 and Table 2 that, as in the case of $\rho^0$, the yields
of $\eta$ and $\omega$ are strongly suppressed at $x_F<0$ as
compared to those in the forward hemisphere. A similar pattern was
observed earlier in $\nu (\overline{\nu})-Ne$ interactions at
higher energies
\cite{ref2}. \\
The production of $\eta'$ was looked for in the channel
$\eta'\rightarrow \rho^0 \gamma\rightarrow \pi^+\pi^-\gamma$ (with
the branching fraction 29.4\%, including the non-resonant
$\pi^+\pi^-$ background). The effective mass of the
$(\pi^+\pi^-)$-system was restricted to the $\rho^0$ mass range
0.6-0.9 GeV$/c^2$, these boundaries being close to those applied
in other experiments in which the $\eta'\rightarrow \rho^0\gamma$
decay fraction had been measured (see references in \cite{ref13}).
The $\pi^+\pi^-\gamma$ effective mass distributions, corrected for
the $\gamma$ detection efficiency, are plotted in Figure 2 (the
right panels). Rather faint signals for $\eta'$ production are
visible, more significant at $x_F>0$. The estimated corresponding
yields corrected for the decay fraction are presented in Table 2.
Note, that the yield of $\eta'$ at $x_F>0$ is significantly
smaller as compared to that for $\eta$, composing $(31\pm25)$\% of
the latter.

c) {\it Strange resonances}

\noindent The production of $K^*(892)^0$ and
$\overline{K}^*(892)^0$ was looked for in channels $K^+\pi^-$ and
$K^-\pi^+$, respectively. The main problem to separate these
resonances is the large background from pion pairs in which the
kaon hypothesis is applied for one of pions. The background from
\begin{table}[ht]
\caption{The mean multiplicities of light resonances.}
\begin{center}
\begin{tabular}{|l|c|c|c|}
  \hline

resonance&all $x_F$&$x_F > 0$&$x_F <0$
\\ \hline
$\eta$&0.050$\pm$0.044&0.051$\pm$0.027&0.015$\pm$0.030 \\
$\rho^0$&0.054$\pm$0.017&0.042$\pm$0.015&0.013$\pm$0.009
\\
$\rho^+$&0.120$\pm$0.031&0.051$\pm$0.015&0.067$\pm$0.025
\\
$\rho^-$&0.039$\pm$0.015&0.010$\pm$0.008&0.031$\pm$0.013
\\
$\omega$&0.053$\pm$0.017&0.044$\pm$0.012&0.005$\pm$0.013
\\
$\eta'(958)$&0.025$\pm$0.020&0.016$\pm$0.010&0.009$\pm$0.016
\\
$f_0(980)$&0.014$\pm$0.010&0.012$\pm$0.008&0.002$\pm$0.009
\\ \hline
$K^*(892)^0$&0.023$\pm$0.013&0.002$\pm$0.008&0.020$\pm$0.012
\\
$\overline{K}^*(892)^0$&0.015$\pm$0.010&0.001$\pm$0.007&0.016$\pm$0.010
\\
$K^*(892)^+$&0.022$\pm$0.012&0.012$\pm$0.010&--- \\
$K^*(892)^-$&0.006$\pm$0.005&---&--- \\
 \hline
$\phi(1020)$&0.009$\pm$0.006&0.008$\pm$0.003&0.002$\pm$0.004
\\
\hline
\end{tabular}
\end{center}
\end{table}
uncorrelated $(\pi^+\pi^-)$ can be approximated by a smooth curve
(according to eq.(3)), while the background from correlated
$\pi^+\pi^-$, originating from the decay of the same parent
particle, can induce peculiarities in the effective mass spectra
calculated for $(K\pi)$ hypothesis. To subtract the correlated
background from $\eta \rightarrow \pi^+\pi^-\pi^0$,
$\omega\rightarrow \pi^+\pi^-\pi^0$ and $\rho^0\rightarrow
\pi^+\pi^-$ decays, we calculated, using their yields extracted in
this work, their contributions to the $(\pi^+\pi^-)$ mass
spectrum. Then each $(K\pi)$ combination was weighted, depending
on the mass of the corresponding $(\pi^+\pi^-)$ pair. In a similar
way, we took also into account the contribution from the
non-identified $K_s^0\rightarrow \pi^+\pi^-$ decays at very close
distances from the neutrino interaction vertex (the magnitude of
these 'close' decays was estimated from
the analysis of $(\pi^+\pi^-)$ mass spectrum). \\
The effective mass spectra for $(K^+\pi^-)$ and $(K^-\pi^+)$,
corrected as described above, are plotted in Figure 3, while the
estimated yields of $K^*(892)^0$ and $\overline{K}^*(892)^0$
(corrected for the decay fractions) are presented in Table 2.
Despite of large relative errors, one can deduce that the yield of
these resonances at $x_F>0$ is suppressed as compared to that at
$x_F<0$. This can be a direct consequence of their valence quark
composition, $K^{*0}(d\overline{s})$ and
$\overline{K}^{*0}(\overline{d}s)$. As it can be estimated from
the parton model (following \cite{ref14,ref15,ref16}), the
probability of creation of favorable $K^{*0}$ and
$\overline{K}^{*0}$ initiated by subprocesses
$\nu\overline{u}\rightarrow\mu^-\overline{s}$ and
$\nu\overline{u}\rightarrow\mu^-\overline{d}$, respectively, and
carrying relatively large positive $x_F$'s, is much smaller, than
that for unfavorable ones produced in the region of $x_F<0$, as a
result of the fragmentation of the quark
string formed in the main subprocess $\nu d\rightarrow\mu^-u$. \\
The production of $K^*(892)^{\pm}$ was looked for in channels
$K_s^0\pi^{\pm}$. The effective mass distributions of the
($K_s^0\pi^+$) and ($K_s^0\pi^-$) systems for the full $x_F$-range
and of the ($K_s^0\pi^+$) system at $x_F > 0$ are plotted in
Figure 4 (the left panels). The shape of the background
distributions was determined with the help of the mixed event
technics, combining $K_s^0$'s with $\pi$ mesons from other events
in which another $K^0_s$ was detected. The parameters of the
background distribution were fixed from the fit by eq. (3), except
the normalization parameter $B$ which was considered as a free
parameter when fitting the experimental distribution by eq. (2).
The fit results for the mean multiplicities (corrected for the
decay fractions) are presented in Table 2. Due to the lack of the
$K^0_s$ statistics, no estimations for yields can be inferred
separately at $x_F > 0$ and $x_F < 0$. \\
If one assumes, that the ratio $R_V(S/N)$ of the summary total
yields of strange $(K^{*0}$, $\overline{K}^{*0}$, $K^{*+}$,
$K^{*-})$ and non-strange $(\rho^0, \rho^+, \rho^-, \omega)$
vector mesons is not significantly influenced by the contribution
from the higher-mass resonances decaying into these mesons, the
$R_V(S/N)$ can be considered as an approximate measure of the
strangeness suppression in the quark string fragmentation process.
Using the data quoted in Table 2, one obtains
$R_V(S/N)=0.25\pm0.09$. This estimate is compatible with the
values of the strangeness suppression factor $\lambda_s$ inferred
from different experiments using various methods (see
\cite{ref17,ref18,ref19} for reviews, as well as the recent works
\cite{ref5} and \cite{ref20}).

d) {\it $\phi$(1020) meson}

\noindent The $(K^+K^-)$ effective mass distribution for different
ranges of $x_F$ are plotted in Figure 4 (the right panel). The
contamination from correlated $(\pi^+\pi^-)$ pair was subtracted
as described in the previous subsection. The $\phi$ yields
corrected for the decay fraction are presented in Table 2. As it
is seen, the production of the $\phi$ mesons occurs, in contrast
to the open strangeness $K^*(892)^0$ and $\overline{K}^*(892)^0$
mesons, only in the forward hemisphere. This can happen if the
$\phi$ meson is a decay product of a favorable meson carrying a
significant fraction of the hadronic energy. The most probable
candidate is the charmed, strange meson $D^+_s$ created directly
as a result of the subprocess $\nu s\rightarrow \mu^- c$ followed
by the recombination of the current charmed quark with the strange
antiquark from the nucleon remnant, or created indirectly as a
result of the decay of the favorable $D^{*+}_s$ meson,
$D^{*+}_s\rightarrow D^+_s \gamma$. The later process (for the
case of $D^{*-}_s$) was observed, with a sufficiently large
probability (about 5\%), in charged current $\overline{\nu}-Ne$
interactions \cite{ref21}. The results of a more detailed analysis
of our experimental data on the $\phi$ neutrinoproduction will be
presented elsewhere \cite{ref22}.

e) {\it The $W$-dependence of the resonance yields}

\noindent Our data on $\eta, \rho$ and $\omega$ neutrinoproduction
combined with those obtained in $\nu(\overline{\nu})-Ne$
interactions at higher energies \cite{ref2}, as well as the data
on $K^*(892)^+$ neutrinoproduction (\cite{ref6} and references
therein) allow one to trace the $W$-dependence of the yields of
these resonances plotted in Figure 5. These dependences can be
approximately described by a simplest linear form $b \cdot
(\langle
W\rangle - W_0)$ at the fixed threshold value $W_0$ = 1.8 GeV. \\
The fitted slope parameters $b$ are given in Table 3. It is
interesting to note that the parameters $b$ for the neutral and
charged favorable $\rho$ mesons coincide at $x_F > 0$ and
significantly exceed those for $\eta$ and $\omega$ which, in their
turn, are almost equal.

\begin{table}[ht]
\caption{The slope parameter $b$ (in GeV$^{-1}$).}
\begin{center}
\begin{tabular}{|l|c|c|c|}
  \hline

reaction&all $x_F$&$x_F > 0$&$x_F <0$
\\ \hline
$\nu(\overline{\nu}\rightarrow)\eta$&0.044$\pm$0.011&0.034$\pm$0.005&---\\
$\nu(\overline{\nu}\rightarrow)\rho^0$&0.056$\pm$0.007
&0.045$\pm$0.004&0.013$\pm$0.005
\\
$\nu\rightarrow\rho^+$ or
$\overline{\nu}\rightarrow\rho^-$&0.091$\pm$0.011&0.046$\pm$0.006
&0.037$\pm$0.006
\\
$\nu\rightarrow\rho^-$ or
$\overline{\nu}\rightarrow\rho^+$&0.023$\pm$0.008&---&---
\\
$\nu(\overline{\nu})\rightarrow
\omega$&0.054$\pm$0.007&0.029$\pm$0.007 &---
\\
$\nu \rightarrow$ $K^*(892)^+$&0.017$\pm$0.002&---&---
\\
\hline
\end{tabular}
\end{center}
\end{table}

f) {\it The mass dependence of the yields of favorable resonances
at $x_F > 0$}

\noindent One can expect, that the main part of the light meson
resonances in the forward hemisphere is a direct product of the
quark string fragmentation, with a comparatively small
contribution from the decay of higher-mass resonances.
Furthermore, their yields in the forward hemisphere are less
influenced by the intranuclear secondary interactions, in
particular by interactions of pions leading to an additional
production of resonances at $x_F < 0$ (see \cite{ref4} for the
case of $\rho^0$ meson). Hence the data on the mass dependence of
the resonance yields at $x_F > 0$ provide an almost
non-deteriorated information on the dynamics of the quark string
fragmentation. The yields of favorable resonances at $x_F > 0$,
normalized to the spin factor $(2J+1)$, as a function of the
resonance mass $m_R$ are plotted in Figure 6 (the left panel),
together with the data of \cite{ref2} which include also the
tensor $f_2(1270)$ meson. Note, that the scaling to the spin
factor is introduced in view of rather small spin alignment
effects in the production of vector mesons (\cite{ref7} and
references therein). As it is seen from Figure 6, the mass
dependence of the resonance yields can be approximately described
by a simple exponential form $A \exp(-\gamma m_R)$. with the
fitted values of the slope parameter $\gamma$ (quoted in Figure 6)
not exhibiting any significant dependence on the initial internal
energy of the quark string. It might be interesting to note, that
a steeper mass dependence was observed in higher energy $e^+e^-$
annihilation (see \cite{ref13} for the data review and
\cite{ref20} for the analysis of the mass dependence).

\section{The fraction of pions originating from the decay \\of light resonances}

\noindent The data presented in the previous section allow one to
estimate the fractions of $\pi^0, \pi^-$ and $\pi^+$ originating
from the decay of light resonances. These fractions were
calculated using the corresponding branching ratios \cite{ref13}.
To avoid double account, the contributions from the indirect
$\eta, \rho, \omega$ (from the $\eta'$ and $\phi$ decays) were not
accounted for. The total yield of pions were taken from
\cite{ref5}. The calculated individual and summary fractions of
decay pions are collected in Table 4. As it is seen, the fractions
of decay $\pi^0$ and $\pi^-$ are compatible, consisting about 1/3,
while
that for $\pi^+$ is significantly smaller, about 1/5. \\
The $W$-dependence of the decay fraction can be obtained for pions
from the decay of the {\it lightest} non-strange resonances
$(\eta, \rho, \omega)$ which provide the main contribution to the
decay pions (cf. Table 4) and for which the data at higher
energies are available \cite{ref2}. This dependence is plotted in
Figure 7, where the corresponding fractions in $e^+e^-$ at 91 GeV
(see \cite{ref13} for references) are also shown for comparison.
As it is seen, the decay fractions continuously increase with
energy.

\begin{table}[th]
\caption{The mean multiplicities of pions (1st row) and resonances
(1st column) and the fraction of pions originating from the
resonance decays (in \%). To avoid double account, the
contribution of $\eta$, $\rho$, $\omega$ from decays of $\eta^{'}$
and $\phi$ is not included.}
\begin{center}
\begin{tabular}{|l|c|c|c|c|}
  \hline

\multicolumn{2}{|c|}{}&$\pi^0$&$\pi^-$&$\pi^+$
\\
\multicolumn{2}{|c|} {total} & \, \,
0.904$\pm$0.066&0.652$\pm$0.010&1.55$\pm$0.06
\\  \hline
\multicolumn{2}{|c|} {$\eta$  $\, \, \, \, \,  ~~~~~~~$
0.050$\pm$0.044}&~~6.65$\pm$5.85&2.10$\pm$1.85&0.88$\pm$0.78
\\  \hline
\multicolumn{2}{|c|} {$\rho^0$ $ \, \, \, \, \,  ~~~~~~~$
0.054$\pm$0.017}&--&8.28$\pm$2.58&3.50$\pm$1.09
\\  \hline
\multicolumn{2}{|c|} {$\rho^+$  $  \, \, \, \, \, ~~~~~~~$
0.120$\pm$0.031}&13.27$\pm$3.56&--&7.74$\pm$2.02
\\  \hline
\multicolumn{2}{|c|} {$\rho^-$  $  \, \, \, \, \, ~~~~~~~$
0.039$\pm$0.015}&~~4.41$\pm$1.73&5.98$\pm$2.30&--
\\  \hline
\multicolumn{2}{|c|} {$\omega$  $ \, \, \, \, \,  ~~~~~~~$
0.053$\pm$0.017}&~~5.75$\pm$1.84&7.38$\pm$2.37&3.10$\pm$1.00
\\  \hline
\multicolumn{2}{|c|} {$\eta^{'}(958)$  $\, \,~~$
0.025$\pm$0.020}&~~1.13$\pm$0.90&2.78$\pm$2.22&1.17$\pm$0.93
\\  \hline
\multicolumn{2}{|c|} {$K^*(892)^0$ $\,$
0.023$\pm$0.013}&~~0.85$\pm$0.49&2.18$\pm$1.26&--
\\  \hline
\multicolumn{2}{|c|} {$\overline{K}^*(892)^0$ $\,$
0.015$\pm$0.010}&~~0.57$\pm$0.38&-- &0.66$\pm$0.45
\\  \hline
\multicolumn{2}{|c|} {$K^*(892)^+$ $\,$
0.022$\pm$0.012}&~~0.81$\pm$0.47&-- &0.95$\pm$0.52
\\  \hline
\multicolumn{2}{|c|} {$K^*(892)^-$
0.006$\pm$0.005}&~~0.20$\pm$0.17& 0.56$\pm$0.47&--
\\ \hline
\multicolumn{2}{|c|}{$f_0(980)$ \, \,
0.014$\pm$0.010}&~~1.03$\pm$0.73&1.42$\pm$1.00&0.60$\pm$0.42
\\  \hline
\multicolumn{2}{|c|}{$\phi$ $\, \, \, \, \, \,  ~~~~~~~$
0.009$\pm$0.006}&~~0.05$\pm$0.03&0.06$\pm$0.05&0.03$\pm$0.02
\\  \hline
\multicolumn{2}{|c|} {\bf{sum}} & \, \,
{\bf{34.7$\pm$7.6\%}}&{\bf{31.2$\pm$5.7\%}}&{\bf{18.4$\pm$3.0\%}}
\\  \hline

\end{tabular}
\end{center}
\end{table}

\section{Summary}

\noindent The total yields of light meson resonances (up to mass
$\sim$ 1 GeV$/c^2$, including $\phi$ meson) are measured in
neutrinonuclear reactions at $\langle E_\nu \rangle \approx$ 10
GeV ($\langle W \rangle$ = 2.8 GeV). For some resonances, the
differential yields in the forward $(x_F > 0)$ and backward ($x_F
< 0$) hemispheres in the hadronic c.m.s. are also obtained. The
data on the inclusive $\phi$ neutrinoproduction are presented for
the first time. It has been shown that the production of favorable
resonances (which can contain the current quark) occurs
predominantly in the forward hemisphere, except for $\rho^+$ meson
the yield of which at $x_F<0$ slightly exceeds that at $x_F>0$. On
the contrary, the production of unfavorable resonances occurs
mainly at $x_F<0$. An exception is $\phi$ meson the production of
which occurs practically only in the forward hemisphere. \\
It is shown from the combined analysis of the obtained and
existing data, that the yields of resonances increase with
$\langle W\rangle$ approximately linearly, in the range of
$\langle W\rangle = 2.8\div4.8$ GeV. \\
The yields of favorable resonances at $x_F > 0$, normalized to the
spin factor $(2J+1)$, are found to be exponentially falling as a
mass function. \\
The fractions of $\pi^0, \pi^-$ and $\pi^+$ originating from the
light resonance decays are estimated to be 34.7$\pm$7.6,
31.2$\pm$5.7 and 18.4$\pm$3.0\%, respectively.

{\bf Acknowledgement.} The authors from YerPhI acknowledge the
supporting grants of Calouste Gulbenkian Foundation and Swiss
Fonds "Kidagan". The activity of one of the authors (H.G.) is
supported by Cooperation Agreement between DESY and YerPhI signed
on December 6, 2002.



\newpage
\begin{figure}[ht]
\centering
 \resizebox{1.0\textwidth}{!}{\includegraphics*[bb =2 20 490 560]
{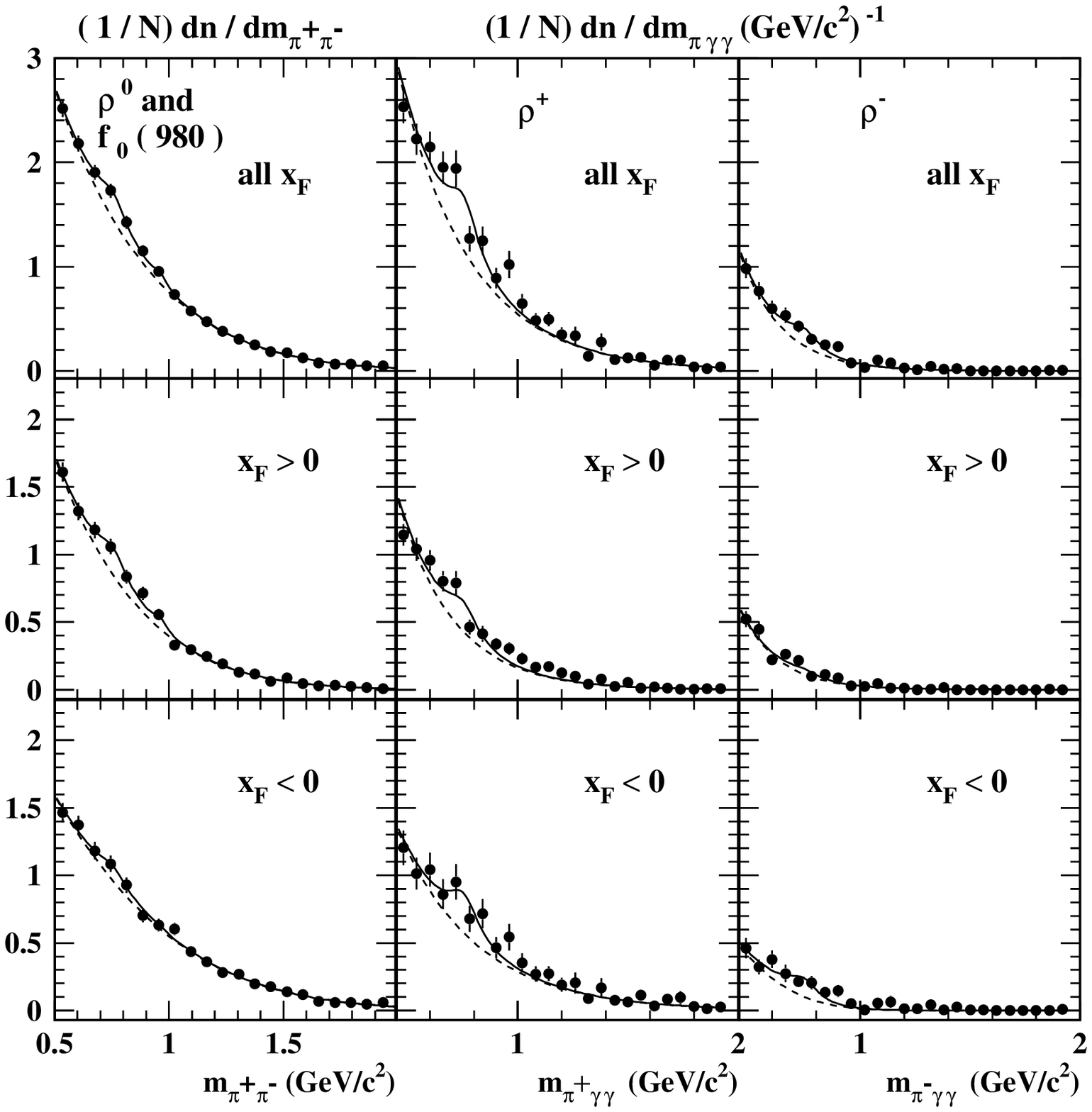}} \caption{The effective mass distributions for systems
$\pi^+ \pi^-$, $\pi^+ \gamma \gamma$ and $\pi^- \gamma \gamma$.
The solid and dashed curves are the fit results for experimental
and background distributions.}
\end{figure}

\newpage
\begin{figure}[ht]
\centering \resizebox{1.3\textwidth}{!}{\includegraphics*[bb=5 40
650 590] {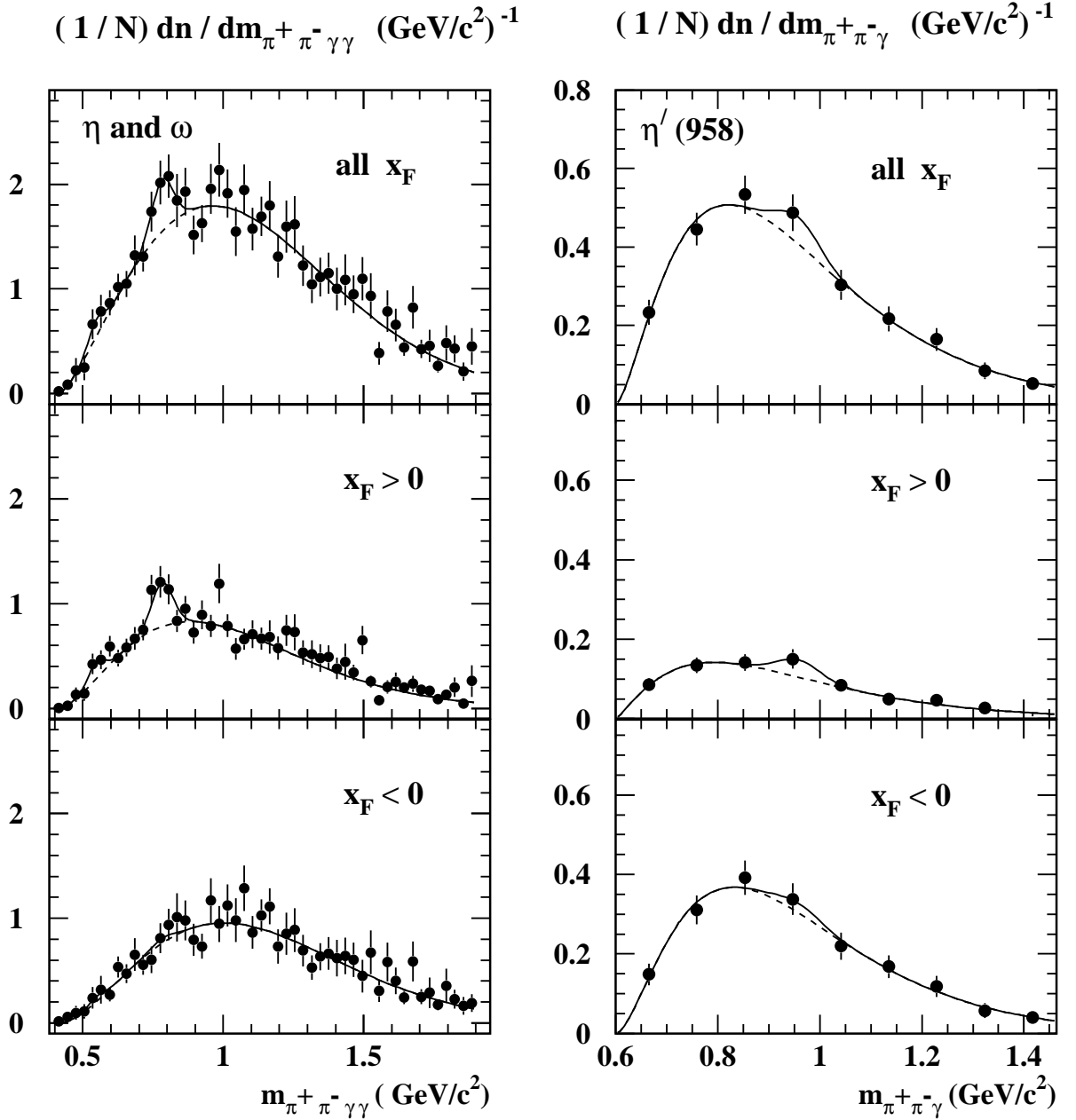}} \caption{The effective mass distributions for
systems $\pi^+ \pi^-{\gamma \gamma}$ and $\pi^+\pi^-{\gamma}$. The
explanation of curves is the same as for Figure 1.}
\end{figure}

\newpage
\begin{figure}[ht]
\centering \resizebox{1.1 \textwidth}{!}{\includegraphics*[bb=20
30 600 590]{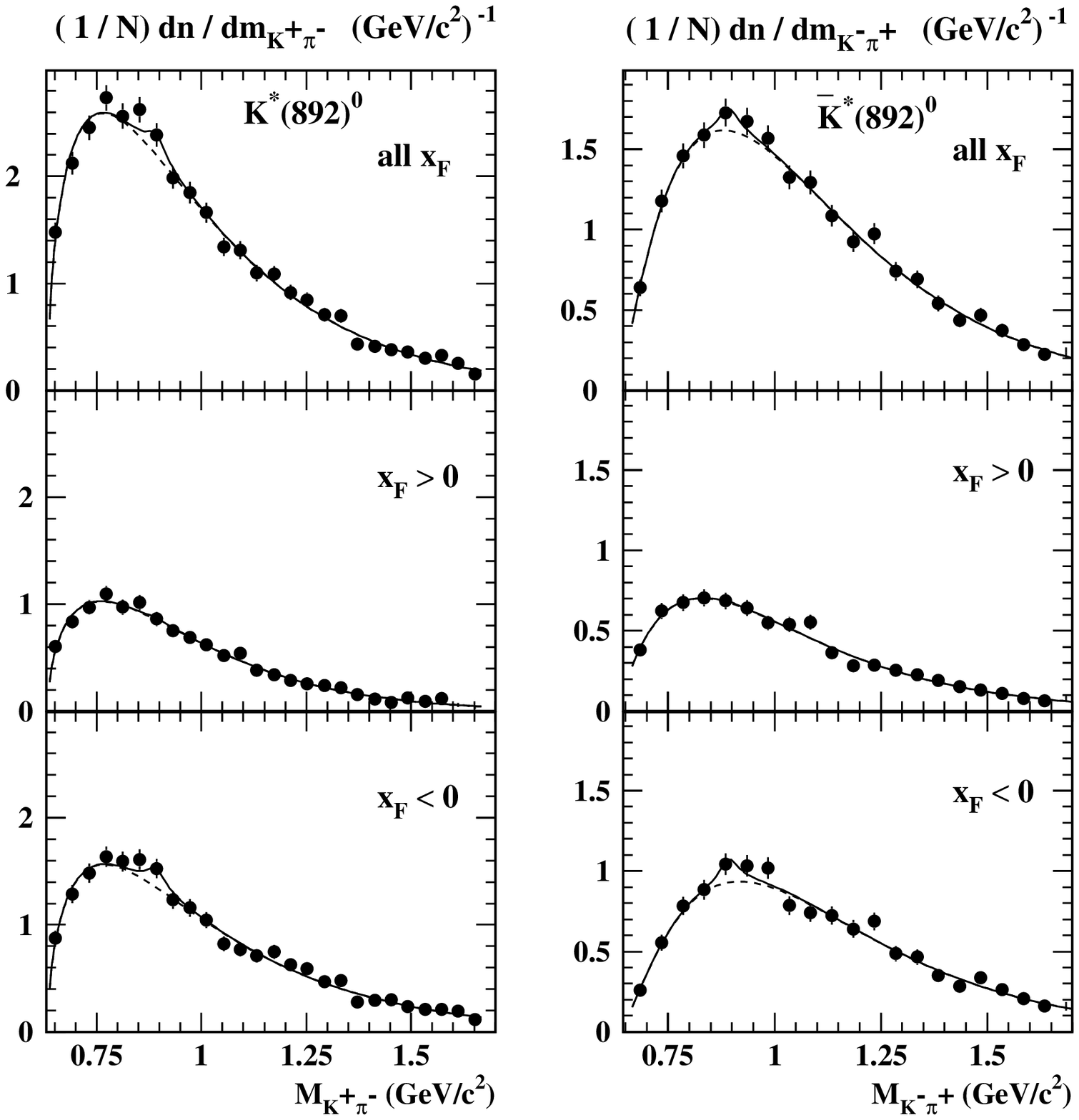}} \caption{The effective mass distributions
for systems $K^+ {\pi^-}$ and $K^-{\pi^+}$. The explanation of
curves is the same as for Figure 1.}
\end{figure}

\newpage
\begin{figure}[ht]
\resizebox{1.1 \textwidth}{!}{\includegraphics*[bb=20 30 600
560]{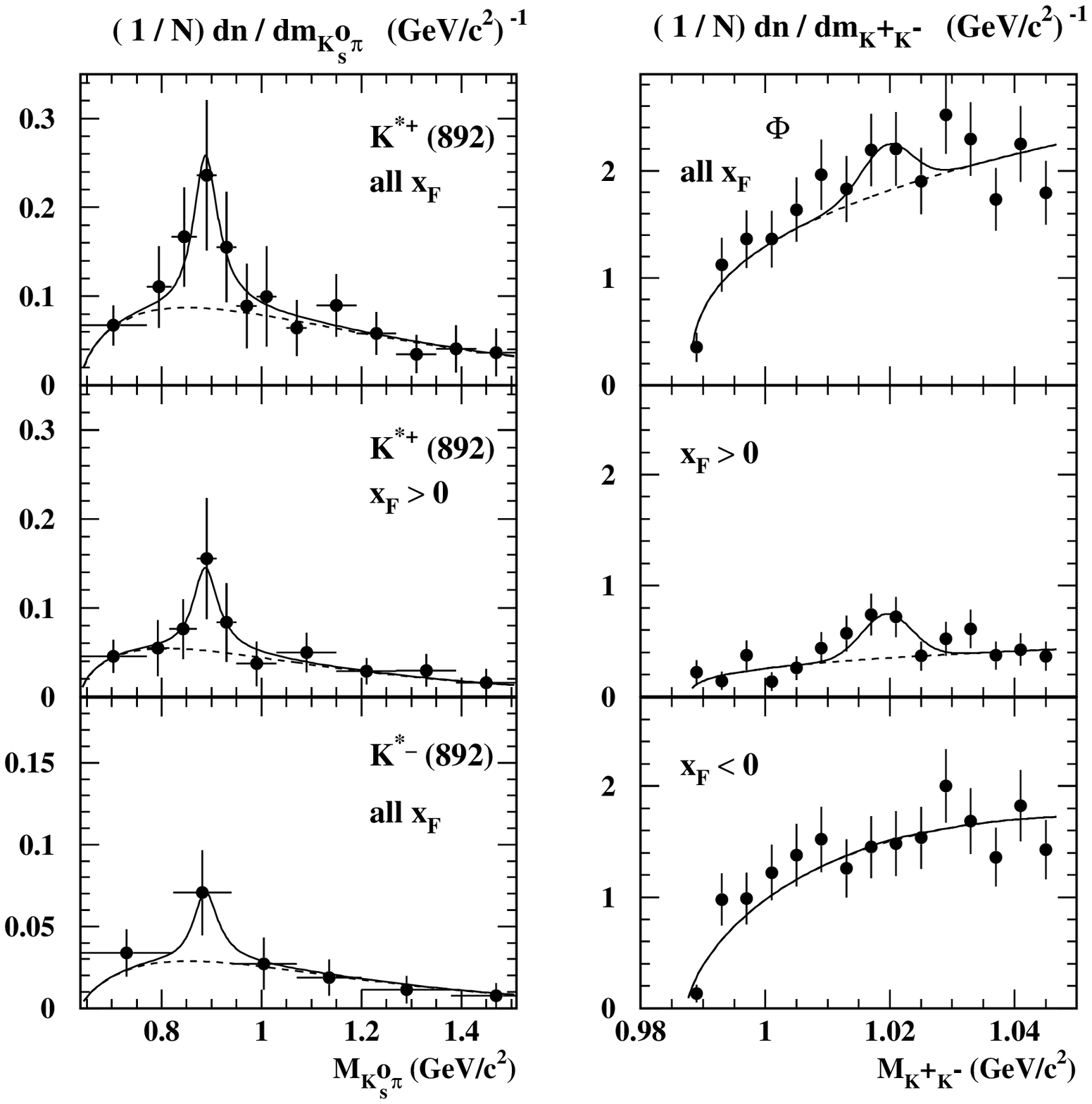}} \caption{The effective mass distributions for
systems $K_s^0 \pi^+$, $K_s^0 \pi^-$ and $K^+ K^-$. The
explanation of curves is the same as for Figure 1.}
\end{figure}

\newpage
\begin{figure}[ht]
\resizebox{1.1 \textwidth}{!}{\includegraphics*[bb=20 30 600
490]{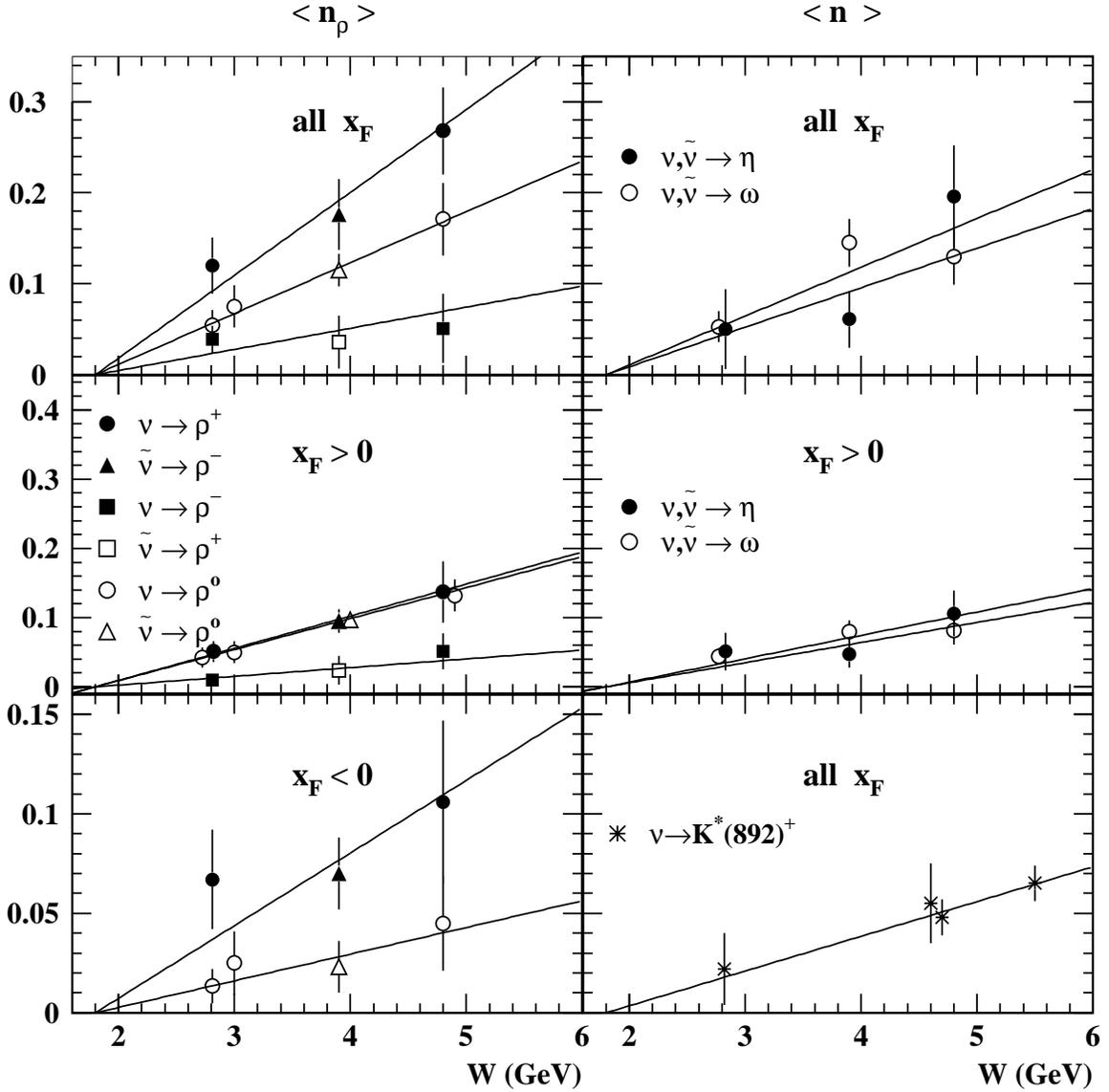}} \caption{The $W$ - dependence of the yields of
$\rho^0$, $\rho^+$ and $\rho^-$ (left), $\eta$ and $\omega$
(right: the top and middle panels) and $K^*(892)^+$ (right: the
bottom panel). The lines are the fit results (see the text).}
\end{figure}

\newpage
\begin{figure}[ht]
\resizebox{1.2\textwidth}{!}{\includegraphics*[bb=20 30 650 550]
{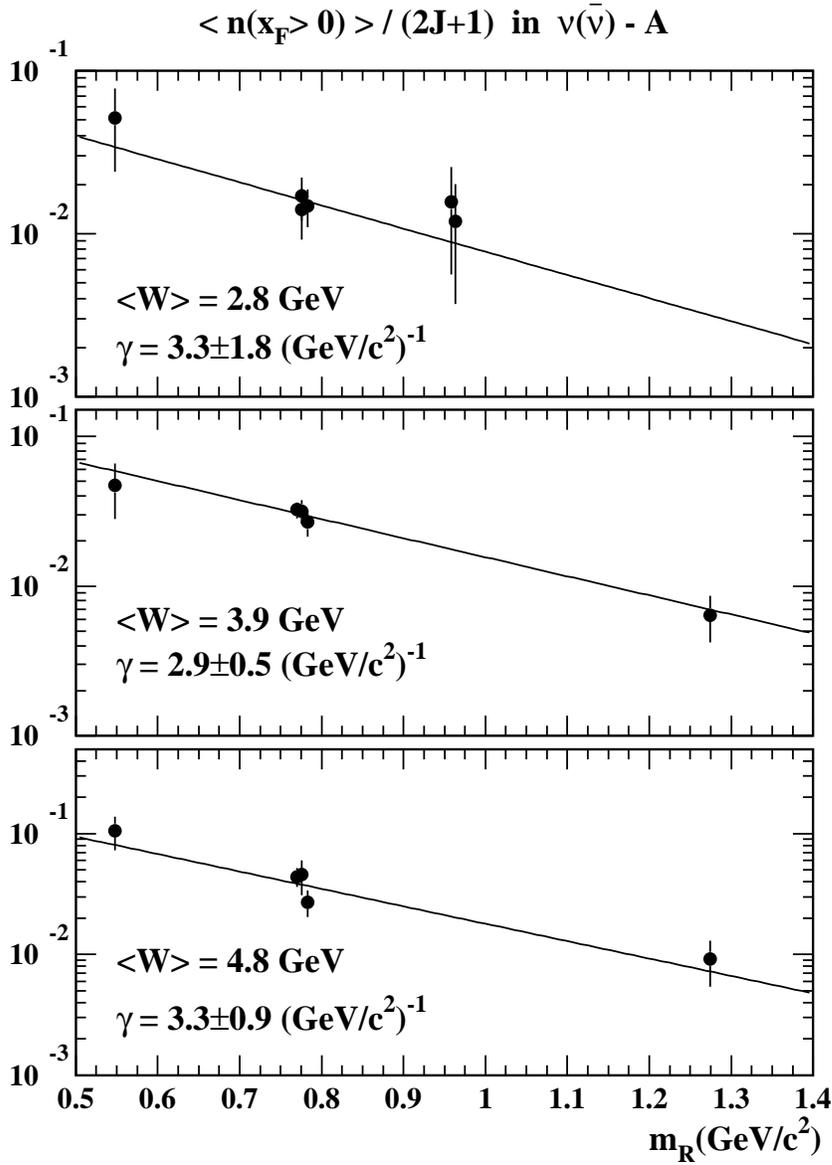}} \caption{The mass dependence of the resonance yields,
normalized to the spin factor $(2J+1)$, in the forward hemisphere
in $\nu(\overline{\nu})-A$ interactions. The lines are the fit
results (see text).}
\end{figure}

\newpage
\begin{figure}[ht]
\resizebox{1.3\textwidth}{!}{\includegraphics*[bb=2 20 650 550]
{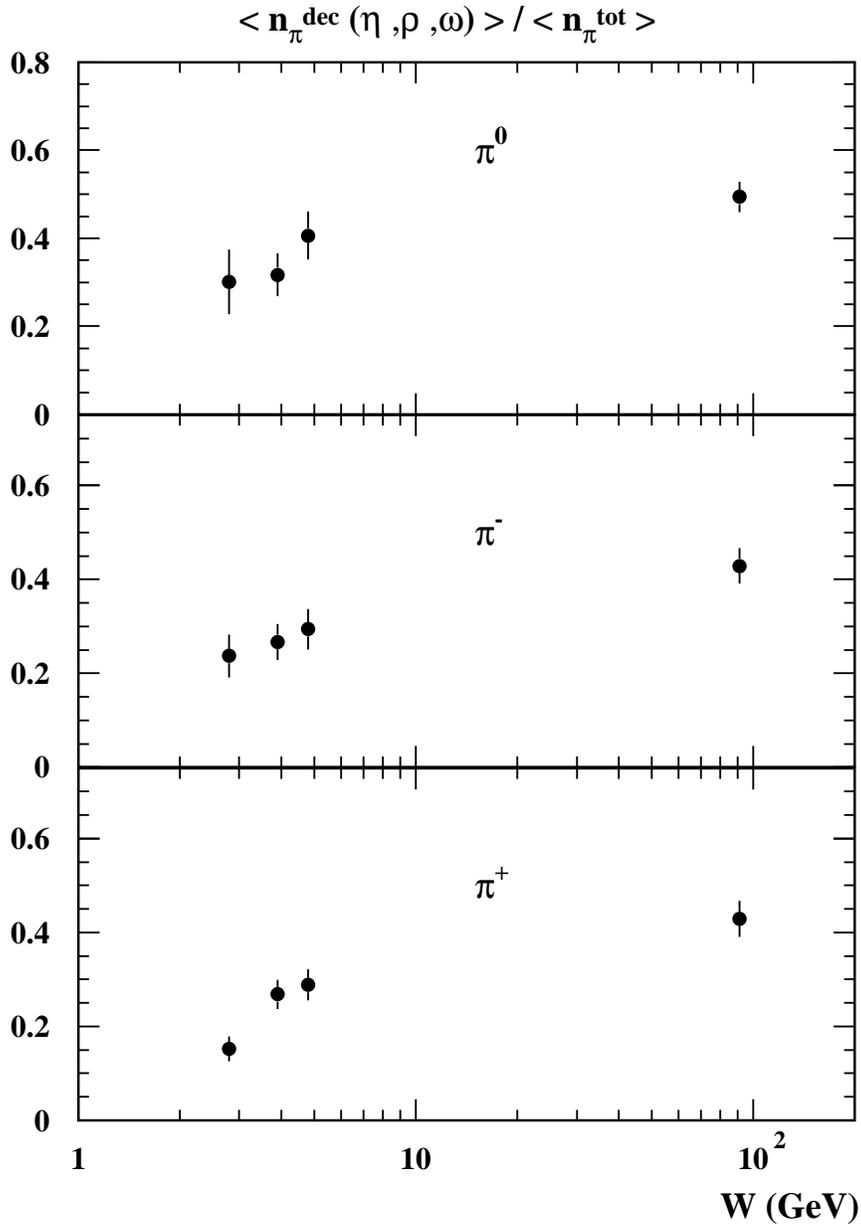}} \caption{The fraction of pions originating from the
$\eta, \rho, \omega$ decays. The data at $W$ = 3.9 GeV for
$\pi^+(\pi^-)$ in $\overline{\nu}Ne$ interactions \cite{ref2} are
replaced by those for $\pi^-(\pi^+)$.}
\end{figure}

\end{document}